\documentclass[a4paper,10pt,twocolumn]{article}
\usepackage[utf8]{inputenc}
\usepackage[T1]{fontenc}
\usepackage{amsmath}
\usepackage{amsfonts}
\usepackage{amssymb}
\usepackage{authblk}
\usepackage{fullpage}

\usepackage{multicol}

\usepackage{ifpdf}
\usepackage{url}

\ifpdf
 \usepackage[pdftex]{graphicx}
 \usepackage[pdftex]{color}
\else
 \usepackage{graphicx}
 \usepackage{color}
\fi
\usepackage{wrapfig}
\usepackage{verbatim}
\usepackage{tikz-cd}

\usepackage[colorlinks=true,urlcolor=blue,citecolor=black]{hyperref}

\usepackage{tcolorbox}
\definecolor{aliceblue}{rgb}{0.94, 0.97, 1.0}

\usepackage{scicite}

\usepackage{flushend}

\newcommand{\rem}[1]{}

\begin{document}

\twocolumn[
{\Large \bf \centering{How social feedback processing in the brain shapes collective opinion processes in the era of social media}}\vspace{4pt}\\
{\large by Sven Banisch, Felix Gaisbauer and Eckehard Olbrich} \vspace{4pt}\\
Max Planck Institute for Mathematics in the Sciences, Leipzig, Germany\vspace{-8pt}\\
\line(1,0) {520}
\vspace{20pt}
]

\textbf{What are the mechanisms by which groups with certain opinions gain public voice and force others holding a different view into silence? 
And how does social media play into this? 
Drawing on recent neuro-scientific insights into the processing of social feedback, we develop a theoretical model that allows to address these questions. 
The model captures phenomena described by spiral of silence theory of public opinion, provides a mechanism-based foundation for it, and allows in this way more general insight into how different group structures relate to different regimes of collective opinion expression. 
Even strong majorities can be forced into silence if a minority acts as a cohesive whole. 
The proposed framework 
of social feedback theory (SFT) 
highlights the need for sociological theorising to understand the societal-level implications of findings in social and cognitive neuroscience.
}

A better understanding of the collective processes underlying public opinion expression is crucial for a better understanding of modern society. 
Sociological models drawing on network science \cite{Newman2002random,Newman2006structure,Borgatti2009network} and basic principles of human interaction behaviour \cite{Bechtel2005explanation,Hedstrom2010causal} have already provided useful insight into collective phenomena related to mass mobilisation \cite{Granovetter1988threshold,Kuran1989sparks,Lohmann2000collective}, societal-level change of behaviour \cite{Centola2011experimental,Bond2012million,Christakis2013social} and beliefs \cite{Friedkin2016network}. 
But for change to happen, for a movement to gain pace, the alternative must be voiced by a sufficiently large group \cite{Centola2018experimental}. 
And to be voiced, it must be perceived as something that can be said without >>fear of isolation<< \cite{Locke1860essay}.


Spiral of silence theory \cite{NoelleNeumann1974spiral} is based on this old >>law of opinion<< \cite[John Locke]{Locke1860essay}. 
It focusses on the collective perception of what can be publicly voiced and hence impact public opinion. 
Noelle-Neumann assumes that humans possess a 
>>quasi-statistical organ<< \cite{NoelleNeumann1996offentliche} to perceive what can be said without being socially sanctioned. 
She frequently refers to the >>social nature of man<< \cite{NoelleNeumann2004spiral} to explain public opinion dynamics as a spiralling process in which silence may lead to more silence. 
In this paper we propose a mathematical model for this process based on reinforcement learning (RL) by social feedback \cite{Banisch2019opinion}. 
In repeated games played over a network, agents receive signals of approval or disapproval on expressing their opinion to peers. 
They evolve an expectation about the social reward obtained when expressing their opinion and remain silent if they expect punishment (negative reward).
RL naturally captures the assumed >>quasi-statistical<< opinion perception. 
Our paper develops a mechanism-based approach \cite{Hedstrom2010causal} that enables a more general application of the assumptions underlying spiral of silence theory and allows to relate structural variations across different groups to different regimes of collective opinion expression.

Moreover, we show that these modelling choices are well-grounded in recent neuro-scientific insights into human social nature.
While the potential for explaining collective behaviour based on mechanisms identified in cognitive and social neuroscience is frequently emphasized \cite{Fareri2014social,Ruff2014neurobiology,Orr2019multi}, its integration with sociological theories of collective opinion expression \cite{NoelleNeumann1974spiral,Granovetter1988threshold,Kuran1989sparks} is lacking.
Social Feedback Theory (SFT) bridges this gap by formulating collective processes of opinion expression as a multi-agent problem in which individual agents adapt according to a reward- and value-based learning scheme identified in neuro-scientific research 
\cite{Schultz1997neural,Haber2010reward,Behrens2009computation,Maia2009reinforcement,Doherty2003temporal,Averbeck2017motivational}.
SFT provides a coherent framework for modelling collective opinion processes that integrates recent neuro-scientific findings, adaptive decision making \cite{Simon1978rationality} and political theory of public opinion \cite{NoelleNeumann1974spiral,NoelleNeumann1996offentliche}.
Such a theoretical framework is essential for the analysis of societal-level implications of social neuroscience.

Social neuroscience aims 
to identify neural mechanisms involved into the processing of social cues. 
to understand their functional role in social cognition.
fMRI studies have shed light on the interaction and interconnectedness of different brain regions and their functional role in social cognition.
While it has long been controversial if human nature evolved a neural circuity specifically for handling social information or not \cite{Poldrack2006can,Dunbar2007evolution,Behrens2009computation,Ruff2014neurobiology}, it is now relatively settled that a basic >>reinforcement circuit<< \cite{Haber2010reward,Fareri2012effects} 
is strongly involved into value-based decisions and learning from social feedback \cite{Izuma2008processing,Klucharev2009reinforcement,Campbell2010opinion,Izuma2010processing,Sherman2016power,Sherman2018brain}.
Other brain processes interfere with this circuity \cite{Haber2010reward,Ruff2014neurobiology,Rilling2011neuroscience} especially when social situations and tasks involve higher cognitive functions such as trust \cite{Fareri2012effects}, morality \cite{Cushman2013action} or representations of self and the other \cite{Amodio2006meeting,Izuma2012social}.

Temporal difference reinforcement learning (TDRL) \cite{Sutton1992reinforcement,Sutton2018reinforcement} has provided a useful computational account of the brain mechanisms underlying social reward processing and learning \cite{Behrens2009computation,Maia2009reinforcement,Doherty2003temporal,Averbeck2017motivational}.
In TDRL, a new estimate of the expected value $Q^{t+1}$ associated to an action is a function of the current estimate $Q^t$ and the temporal difference (TD) error $\delta^t$ between this estimate and the reward that is actually obtained:
$
    Q^{t+1}=Q^{t} + \alpha \delta^t
$.
With a rate governed by $\alpha$ (referred to as learning rate) this scheme  converges to a stable equilibrium in which the TD error $\delta^t$ approaches zero such that expectations and actual reception of rewards are aligned \cite{Sutton2018reinforcement}.
The usefulness of TDRL in computational neuroscience 
derives from the finding that the activity of dopaminergic neurons in the midbrain regions is quantitatively related to the >>reward-prediction error<< \cite{Ruff2014neurobiology} between the experienced reward and its expected value \cite{Schultz1997neural,Hollerman1998dopamine,Doherty2004reward}, that is, to $\delta_t$.
Social neuroscience has provided ample evidence that such a basic reward processing circuit is also highly involved into peer influence processes \cite{Campbell2010opinion,Sherman2016power}, social conformity \cite{Klucharev2009reinforcement} and approval \cite{Izuma2010processing}.

Given that the processing of and learning by social feedback is so deeply routed in the human brain, it is 
of uttermost importance to better understand the collective consequences of these processes.
Especially in social media, a tremendous number of quick feedback decisions is made day by day by billions of users.
>>Like buttons<< and quantitative markers of collective endorsement can be associated with low cognitive costs which suggests that a dominant role is played by the fast value processing mechanisms accounted for by TDRL.
Recent studies have provided some evidence for that 
 \cite{Sherman2016power,Sherman2018brain}.
While it is reasonable to assume that the social reward circuit has evolved to facilitate cohesion and cooperation in small groups \cite{Ruff2014neurobiology} with intensive pairbonding \cite{Dunbar2007evolution}, this reasoning may not apply for societies of increased complexity \cite{Dunbar1998social,Andersson2019toward}.
In complex social networks, human ability to coordinate with in-groups may come at the expense of an increasing alienation to out-groups and therefore drive polarization dynamics \cite{Banisch2019opinion}.
Here we show that SFT of opinion expression provides a neuro-biologically grounded explanation of collective processes involved into >>spirals of silence<< \cite{NoelleNeumann1974spiral} and related phenomena of collective opinion expression.

We consider the situation that two groups with different standpoints on a controversial issue have evolved and engage in public discourse. 
Individuals within both opinion groups can decide to express (E) their standpoint or to be silent (S). 
They receive supportive feedback from their respective in-group and negative feedback from agents in the out-group when expressing their opinion.
Individual interaction is hence formulated as repeated opinion expression games with a reward system that captures approval and disapproval by peers:
\begin{equation}
    r_{i}^{t}=\left\{
\begin{array}{lll}
-c &\text{silent neighbor} \\
-c+1& \, \text{agreement} \\
-c-1& \, \text{disagreement}\\
\end{array}
\right.
\end{equation}
The parameter $c$ corresponds to a fixed cost of expression.
Having received a social feedback reward during an interaction, agents update the expected value $Q_i(A)$ of their current action by TDRL 
\begin{equation}
    Q_{i}(A)^{t+1}=Q_{i}(A)^{t} + \alpha \underbrace{( r_{i}^{t} - Q_{i}(A)^{t})}_\text{TD error}
    \label{eq:TDIndividual}
\end{equation}
with learning rate $\alpha$.
As the reward of silence (S) is zero in the game we only have to keep track of the value for opinion expression and skip action indices in the sequel ($Q_i(E) = Q_i$).
Given the current value of opinion expression $Q_i$ an agent has learned in previous interactions, the probability of opinion expression follows a softmax choice model of the form
\begin{equation}
p_i = \frac{1}{1+e^{-\beta Q_i}}
\label{eq:actionselection}
\end{equation}
in which $\beta$ governs the rate of exploration.
Taken together the action selection (\ref{eq:actionselection}) and
the TDRL scheme (\ref{eq:TDIndividual}) naturally account for the effect that agents become more (less) willing to speak out after receiving positive (negative) feedback.

Assume that we can characterise the two groups in terms of their sizes ($N_1$ and $N_2$), their in-group cohesion and inter-group connectivity.
The probability of in-group influence is $q_{11}$ for group 1, and $q_{22}$ for group 2.
Interaction probability across groups is denoted by $q_{12}, q_{21}$ respectively. 
We define the \emph{structural strength} of group 1 and 2 (denoted as $\gamma$ and $\delta$) as
\begin{equation}
\gamma=\frac{(N_{1}-1)}{N_{2}}\frac{q_{11}}{q_{12}} \text{ and }
\delta=\frac{(N_{2}-1)}{N_{1}}\frac{q_{22}}{q_{21}}.
\label{eq:structuralpower}
\end{equation}
The structural strength of a group is determined by the relative size of the group and the relative in-group connectivity or \emph{cohesion} \cite{Wasserman1994social,Morris2000contagion}.
As $\gamma$ and $\delta$ determine the probability of in-group versus out-group interaction ($\gamma/(\gamma+1)$ versus $1/(\gamma+1)$ for group 1), they also govern the expected rewards for opinion expression for individuals in the two groups
\begin{equation}
    \mathbb{E}(r_{1})= p_{1} \frac{\gamma}{\gamma+1}  - p_{2} \frac{1}{\gamma+1} -c ,
\end{equation}
\begin{equation}
    \mathbb{E}(r_{2})= p_{2} \frac{\delta}{\delta+1} -  p_{1} \frac{1}{\delta+1} -c ,
\end{equation}
where the probabilities for opinion expression $p_1, p_2$ are given by (\ref{eq:actionselection}). 
The Q-values are updated by Eq. (\ref{eq:TDIndividual}) substituting the agent index $i$ by the respective group index and the reward with the expected rewards derived above.
As visible in Eq. (\ref{eq:TDIndividual}), in TD learning the change of Q-values from one time step to the other is given by the TD error times the learning rate $\alpha$.
In the continuous time limit\cite{Tuyls2003selection,Sato2003coupled,Kianercy2012dynamics}, we describe the model dynamics by a system of two differential equations
\begin{eqnarray}
        \dot{Q_1} =  \mathbb{E}(r_1) - Q_1 \ \nonumber\\
        \dot{Q_2} =  \mathbb{E}(r_2) - Q_2.
        \label{eq:2Dsystem}
\end{eqnarray}
As the right hand side is zero when the Q-value estimate is equal to the expected reward, the fixed points of (\ref{eq:2Dsystem}) are possible equilibria of the associated collective game.\footnote{See \cite{Gaisbauer2019dynamics} for a game theoretic analysis of the model and further details on the mean field approach.} 

\begin{figure*}[ht]
 \centering
 \includegraphics[width=0.99\linewidth]{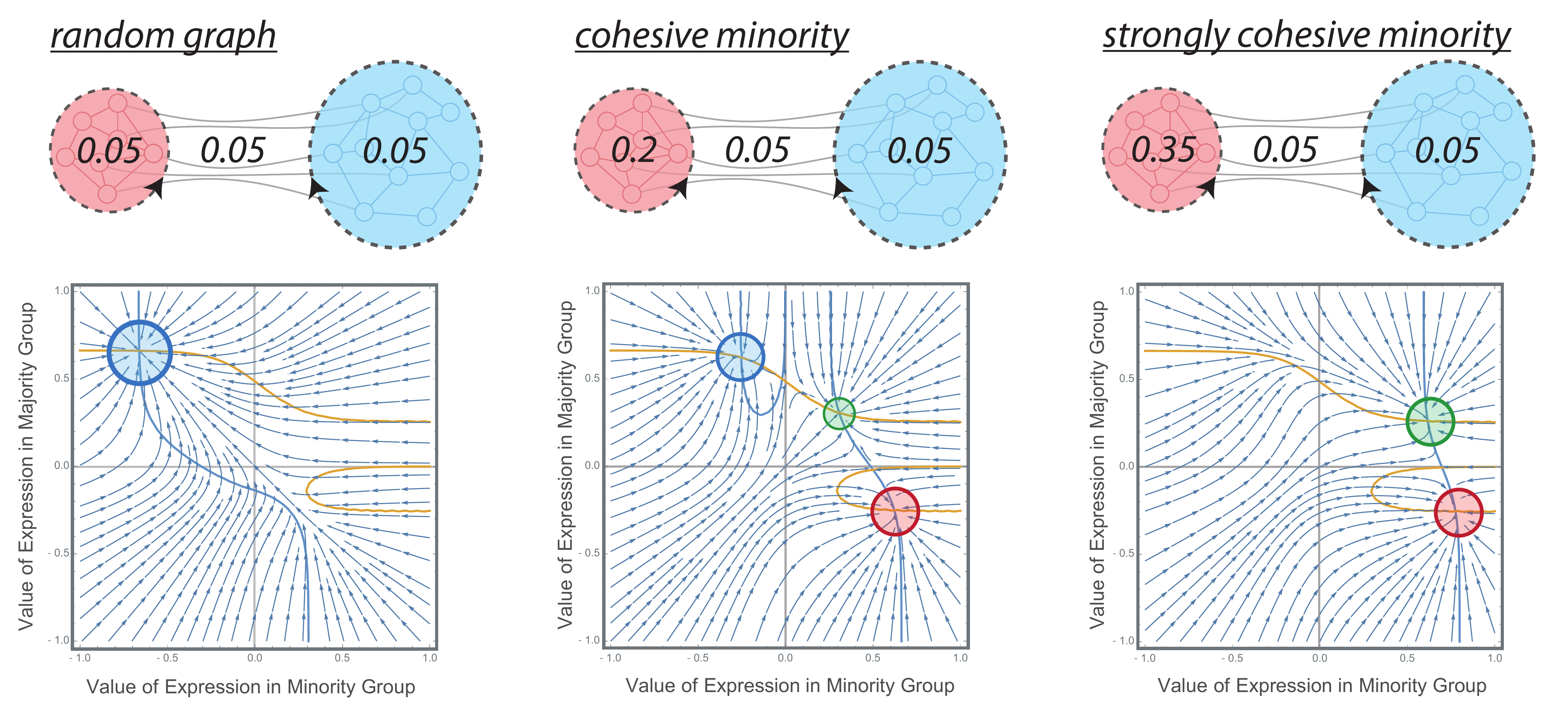}
 \caption{\small Two groups supporting two different opinions struggle for public expression. The majority group (blue) is twice as big as the minority group (red). The three different situations represent subsequent increases of internal cohesion of the minority group and the respective phase dynamics of the system. The isoclines of the dynamics system are shown and the stable fixed points at their intersection are highlighted. While minority expression is unstable in an unstructured random graph, the minority can compensate their quantitative inferiority by a stronger internal organisation. Results for an exploration rate $\beta = 8$.
 Random Graph (left): ER graph with link probability 0.05. For group sizes $N_1 = 100$ and $N_2 = 200$ each agent has 15 links on average. The minority is connected to 5 agents of the in-group and to 10 agents of the out-group. Vice versa for the majority leading to structural strength indicators $\gamma = 0.5$ and $\delta = 2$. The resulting system has only one stable fixed point at $Q_1 \approx -0.33$ and $Q_2 \approx 0.66$ with associated expression rates of $p_1 \approx 0.067$ and $p_2 \approx 0.995$.
That is, only the majority group is expressive and the minority silent. 
  Cohesive Minority (centre): Increasing internal organisation of the minority group by raising the connection probability within the minority to 0.2. This increased group cohesion is reflected in an increased structural strength $\gamma = 2$. $\delta$ is not affected. The system becomes symmetric and minority (red circle) or majority group expression (blue circle) are solutions reached depending on the initial values of expression. An additional fixed point (green circle) emerges in which two groups are loud. 
  Strongly Cohesive Minority (right): The in-group cohesion of the minority further increases leading to $\gamma = 4$ and $\delta = 2$. The case that only the majority is in expression mode is no longer stable and the minority will always express its opinion. Co-existence is still possible.
  }
 \label{fig:ModelResults01}
 \end{figure*}

We apply this model to a minority-majority setting in which one third of the population supports opinion 1 and the other two thirds hold the majority view opinion 2.
The group size ratio $(N_2-1)/N_1$ approaches 2 for a large number of agents.
In the first scenario, the interaction probabilities are homogeneous over the entire population ($q_{11}=q_{22}=q_{12}=q_{21} = q$).
This corresponds to the Erd{\H{o}}s-R{\'e}nyi random graph \cite{ErdHos1960evolution,Gilbert1959random} with link probability $q$ and represents a situation without any particular organisation of social relations within and in between both camps.
The structural strength indicators (\ref{eq:structuralpower}) are then determined by the relative group sizes: $\gamma = 1/2$ and $\delta = 2$. 
The only fixed point of system (\ref{eq:2Dsystem}) corresponds to majority expression and a silent minority (see Figure \ref{fig:ModelResults01}, l.h.s.).
Even if expressive in the beginning, agents in the minority find less and less public support for their opinion and in turn increasingly avoid to express their opinion in public.

\begin{figure}[ht]
 \centering
 \includegraphics[width=0.99\linewidth]{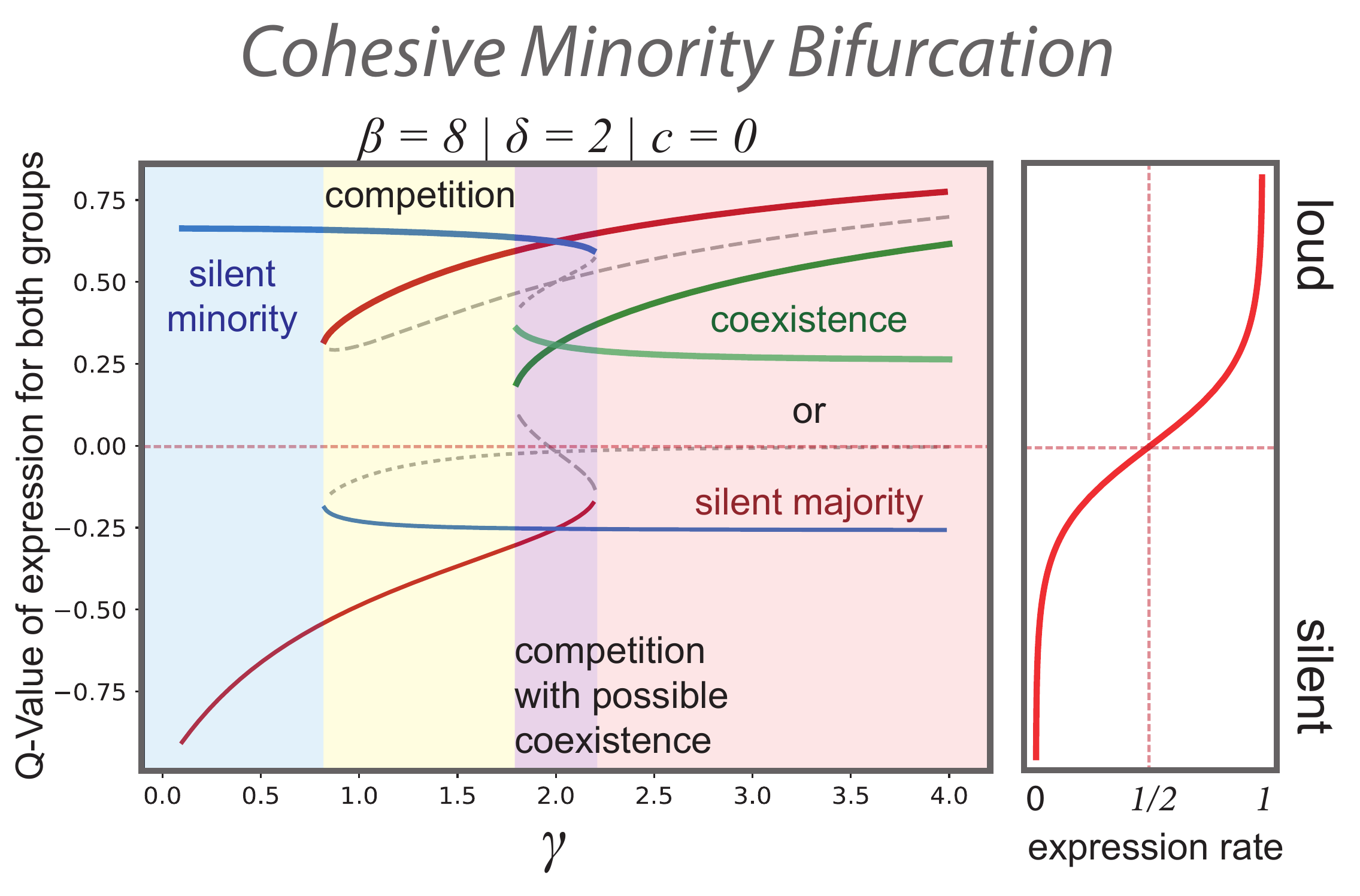}
 \caption{\small Bifurcation plot of the scenario in which a minority ($N_1/N_2 = 1/2$) gains public voice through stronger internal organisation (see also Figure \ref{fig:ModelResults01}). On the r.h.s. the expression rate (\ref{eq:actionselection}) is shown as a function of $Q$ for $\beta = 8$. As the internal cohesion of the minority group increases, for instance, due to strategic linking or tying group symbols, the system undergoes a series of saddle-node bifurcations. Minority expression becomes more and more likely. Solid lines show the Q-values at the stable fixed points. For equilibria in which only one group is loud blue lines represent the majority group $Q_2$, red lines the minority $Q_1$. Green lines correspond to the coexistence equilibrium in which both groups are expressive. While an unorganised minority is forced into silence (blue regime), a slight increase in group cohesion makes minority expression a stable outcome if the majority is silent (competition, yellow). At a certain point ($q_{11} = 4q$ and $\gamma=\delta = 2$) minority organisation can compensate numerical inferiority and stable coexistence of two expressive groups is possible. By further increase of minority group strength, it is always visible in public while the majority may enter a spiral of silence. }
 \label{fig:MinorityBifurcation}
 \end{figure}

The minority can gain public impact if the internal organization of the group becomes more cohesive.
The effect of this structural transition towards stronger minority organisation is shown in Figure \ref{fig:ModelResults01} and \ref{fig:MinorityBifurcation}.
As the probability $q_{11}$ of in-group connection increases the system undergoes a series of saddle-node bifurcations.
First, a small increase of $q_{11}$ (and hence $\gamma$) gives rise to an additional stable fixed point in which only the minority is expressive (not shown).
Minority and majority compete for public voice.
As the internal connectivity of the minority group increases to $q_{11} = 4q$ the situation becomes symmetric with $\gamma = \delta = 2$.
In other words, the minority can compensate quantitative inferiority by a more cohesive internal organisation.
Both groups can readily express their opinion if the other group is silent (competition, Figure \ref{fig:ModelResults01}, centre).
But also an additional stable fixed point in which both opinions coexist appeared through another saddle-node bifurcation (coexistence).
Finally, if the internal cohesion of the minority group becomes very large ($q_{11} = 7 q$), the fixed point associated to a loud majority and silent minority disappears.
That is, the minority always voices their view in public while the majority may become silent (see Figure \ref{fig:ModelResults01}, r.h.s.).

In the model, agents observe and react to their social environment in a way that is strongly reminiscent of Noelle-Neumann’s theory of the spiral of silence \cite{NoelleNeumann1974spiral,NoelleNeumann1996offentliche,NoelleNeumann2004spiral}.
In repeated interaction within their local neighbourhood, agents form 
a >>quasi-statistical<< impression of the current opinion climate
in terms of an internalized expectation ($Q$-values) of which opinion is prevalent in their public spheres and whether their opinion can be articulated without being sanctioned.
If their opinion corresponds to the perceived majority view, they become more willing to speak out. 
If they perceive themselves to hold the minority view, they become less willing to do so. 
If all agents adapt to the current opinion landscape in this way, minorities are forced into a spiralling process in which silence leads to more silence. 
However, only if the minority is perceived as minority in both groups.
Bifurcation analysis of our model shows (see Figure \ref{fig:MinorityBifurcation}) that also majorities can be forced into silence if a minority acts as a cohesive whole.
Even a slight increase of homophily with respect to minority interaction can lead to a situation where a loud minority dominates public discourse because the majority is silent.
Individuals with the actual majority opinion learn that voicing their view in public is rarely answered by support and is more often challenged by an expressive minority.
Silence of the majority group is then collectively reinforced because each individual member is worse off by opinion expression. 

\begin{figure}[ht]
 \centering
 \includegraphics[width=0.99\linewidth]{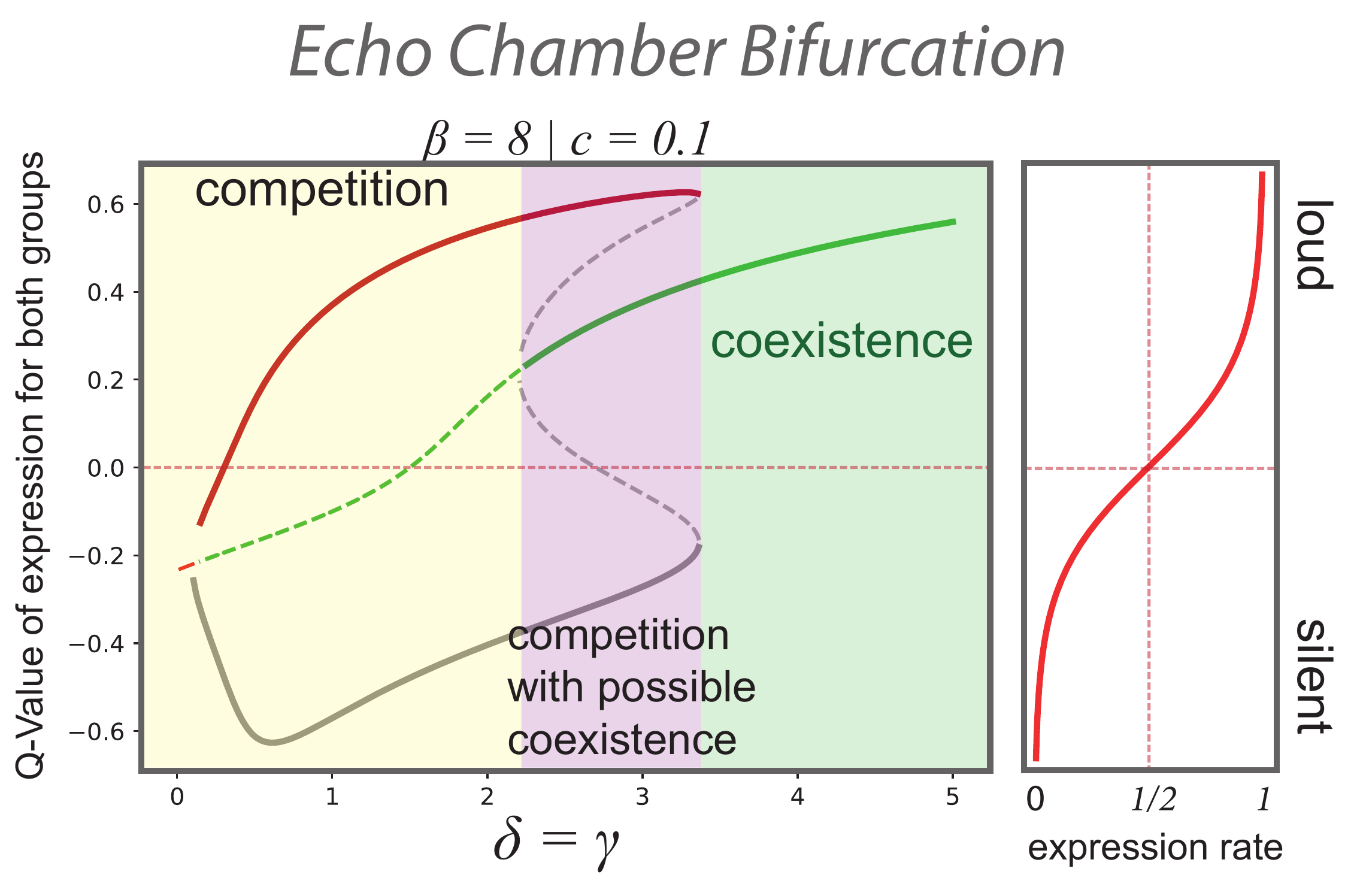}
 \caption{\small 
 Bifurcation plot of the scenario in which two groups of equal size become more structured around the opinion they support. An increase of homophily for both opinion groups is captured by increasing $\gamma$ and $\delta$ at the same time. The situation is symmetric and only $Q_1$ is shown. After a phase of competition if homophily is low (yellow), coexistence emerges as a fixed point (violet) and becomes the only solution after a further slight increase of homophily (green). Both groups express their opinion within their own niches.}
 \label{fig:EchoChamberBifurcation}
 \end{figure}

Spiral of silence theory emerged as an attempt to explain a series of >>last minute swings<< during German elections in the sixties and seventies \cite{NoelleNeumann1996offentliche}. 
Termed >>bandwagon effect<< this phenomena had already been observed by Lazarsfeld and colleagues in the 1940 US presidential elections \cite{Lazarsfeld1944people}. 
While surveyed voting intentions where head-to-head between the two major parties until the very last days of the campaigning period, Noelle-Neumann observed that the evolution of expectations about who will win the election showed a clear trend towards the final winner during month before the election day. 
Developing a series of refined survey instruments, she showed that differences in the willingness to publicly support a party are one source of these trends. 
Our model captures this dynamical feedback between internalized expectations of majority and willingness to actively speak out for one’s party and suggests that the situation of election campaigns at that time is characterised by the competitive regime in Figure \ref{fig:MinorityBifurcation} and~\ref{fig:EchoChamberBifurcation}. 
Our research shows that the collective process described by Noelle Neumann — that is, the spiral of silence — is only one possible outcome of the mircoassumptions on which the theory builds.
The bifurcation analysis of the dynamical system (\ref{eq:2Dsystem}) reveals how structural transformations of group interaction may lead to qualitatively different regimes of collective opinion expression. 

Today, social media are rapidly transforming the landscape of public opinion expression providing niches for virtually every opinion.
As social network services have flexibilized options to connect with like-minded others or to gather under common tags (such as \#FridaysForFuture and \#MeToo) -- and hence to learn that there are others who share a similar view --, previously unseen opportunities to escape the >>fear of isolation<< have grown.
Group interaction that is more and more structured around opinion is captured by the transition shown in Figure \ref{fig:EchoChamberBifurcation}.
As in-group ties become more prevalent in both groups, the system undergoes two saddle node bifurcations from a competitive regime to a regime where coexistence is the only stable outcome.
Private or semi-public rooms for expressing opinion online act as >>echo chambers<< 
and enable that opinions previously marginalized or placed under taboo -- including those advanced by >>populist alternatives<< -- may resist the spiral of silence and become salient in the more general public discourse. 
This alters the public perception of what can be said in public. 
Democratic societies currently struggle with this transformed climate of opinions because the foundational idea of government build on public opinion \cite{Hume1742essays,Lincoln1863gettysburg} is fundamentally challenged.






 \begin{figure*}[ht]
  \centering
  \includegraphics[width=0.98\linewidth]{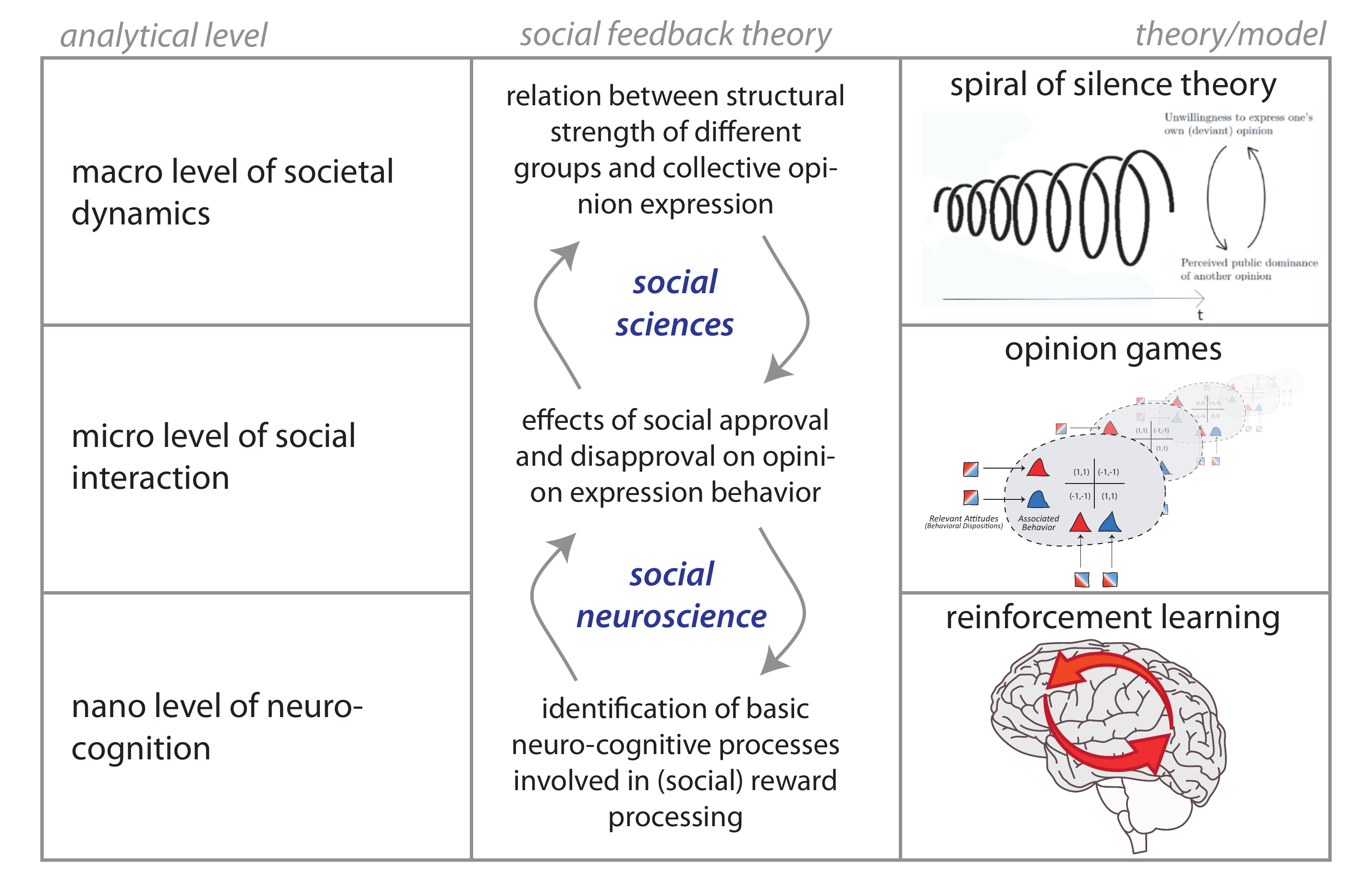}
  \caption{Schematic summary of social feedback theory of opinion expression. The theory involves three analytical levels from the level of neuro-cognitive processes, to the level of social interaction, to the macro level of collective dynamics. SFT bridges these levels through the notion of opinion games: first, by assuming that agent behavior and the associated expected rewards adapt according to a reinforcement learning scheme that resembles the reward processing system in the brain; and second, by showing that agents following this scheme for learning how to act in repeated opinion games leads to collective dynamics as described by spiral of silence theory.
 }
  \label{fig:TableSFT}
  \end{figure*}

SFT aims to contribute to a better understanding of societal-level implications our social nature.
It provides a link between recent research on the neurological basis of social behaviour and sociological theory of public opinion formation and expression.
The model presented in this paper involves abstractions and assumptions at three different analytical levels (see Figure \ref{fig:TableSFT}) each being subject to intensive research from different disciplinary angles.
Providing a coherent theoretical account that integrates sociological modes of structural explanation \cite{Borgatti2009network,Hedstrom2010causal}, adaptive decision theory \cite{Simon1978rationality,Sutton1992reinforcement} and its underlying neurological mechanisms \cite{Dayan2008decision,Glimcher2013neuroeconomics}, SFT offers a unique framework for guiding future research in sociology, decision science and neurobiology.

At the collective level (top row), SFT relates structural transformations in how we interact with one another to different regimes of collective opinion expression.
The main modelling assumption made at this level is to map complex networks of social interaction to the relations within and across groups.
Network science has brought about a portfolio of graph models to more realistically capture social interaction patterns \cite{Scott1988social,Wasserman1994social,Newman2002random,Lee2019homophily} to which the model but not necessarily its formulation as a 2D system of differential equations can be applied.
On the other hand, our theory suggests that social feedback networks generated on the basis of digital trace data \cite{Lazer2009computational} 
are inherently biased by the activity of users who learnt that interaction on the media is rewarding.
In fact, our model suggests that retrieved interaction patterns such as retweet networks \cite{Conover2011political,Garimella2018quantifying,Gaumont2018reconstruction} may render a situation more polarized than it actually is because public expression is less rewarding for actors who maintain relations across different opinion camps.

The micro level of social interaction (second row) is conceived as repeated opinion expression games in which agents respond to one another with signals of approval or disapproval.
This entails simplifications such as dyadic interaction and a reward system that is homogeneous across individuals and groups.
However, by drawing on games in modelling individual interaction, SFT is well-equipped to take into account individual differences in reward perception as well as characteristics of the incentive structure of different platforms.
It shifts the explanatory focus from forms of social influence to the rewards and incentives of opinion expression in different social settings \cite{Banisch2019opinion}.
Opinion games are also flexible enough to include cognitive costs associated to, for instance, preference falsification \cite{Kuran1989sparks,Kuran1991now}.
Seen from the perspective of opinion games, agents rely on TDRL to >>solve<< the associated collective game \cite{Sutton1992reinforcement,Banisch2017coconut,Gaisbauer2019dynamics}.
This enables the application of game theoretic concepts in model analysis \cite{Morris2000contagion,Jackson2014games}.
However, as social neuroscience unequivocally demonstrates, the bounded and procedural account of rationality \cite{Simon1978rationality} implemented by TDRL does not necessarily involve conscious calculation.

The social feedback framework draws on this neuro-cognitive foundation of TDRL (bottom row).
In order to demonstrate that biologically-rooted mechanisms of reward and value processing capture collective processes described in spiral of silence theory, we rely on the most simple TDRL scheme in the model.
Social neuroscience is quickly advancing towards a better understanding how brain areas related to cognitive control interfere with this basic reward circuit. 
Recent work revealed, for instance, that neural responses to social feedback are influenced by the social relation with the interaction partner \cite{Hughes2018wanting} and that the reward valuation circuit is highly involved in shaping these relations \cite{Zerubavel2018neural}.
Experimental designs that mimic interaction on social media \cite{Sherman2016power,Sherman2018brain} could clarify the role of different reward systems and target which kinds of interaction devices activate cognitive control.
In this way, social neuroscience could significantly contribute to a better understanding of the types of games that are actually played in social media environments. 


Considerable progress has been made in social neuroscience in shedding light on human social nature.
Biological evolution has shaped a 
reward processing architecture that is highly involved in social interaction on opinions. 
Social feedback theory allows to address the collective implications that may result from this involvement.
Basal brain mechanisms governing our reactions to social approval and disapproval may have a tremendous impact on collective processes of opinion expression and may be at the root of phenomena such as silent majorities.
Social media facilitates massive and strategic social organisation around opinions. 
This can, as we show, fundamentally alter the perception of public opinion in a society.


 \begin{center}
\line(1,0) {120}
\vspace{12pt}
  \end{center}

\textbf{Acknowledgements.}
This project has received funding from the European Union’s Horizon 2020 research and innovation programme under grant agreement No 732942 (www.\textsc{Odycceus}.eu).
The paper has benefited from many fruitful discussions within Odycceus group seminar in Leipzig. 
We especially acknowledge fruitful interactions with Roger Berger, Marcel Sarkoezi and Armin Pournaki.
We would like to thank Wolfram Barfuss, Marc Keuschnigg and Stefan Westermann for their valuable feedback on earlier versions of this paper.
Thanks also to Thomas Endler for his sketch of the human brain in Figure \ref{fig:TableSFT}.

\small

\end{document}